\documentclass[%
 reprint,
 amsmath,amssymb,
 aps,
]{revtex4-1}

\usepackage{graphicx,youngtab}
\usepackage{dcolumn}
\usepackage{bm,color}

\def\bea{\begin{eqnarray}} 
\def\eea{\end{eqnarray}} 
\def\mC{\mathbb{C}} 

\begin{document}

\preprint{QMUL-PH-17-06}

\title{ Free quantum fields in 4D and Calabi-Yau spaces.  }

\author{Robert de Mello Koch$^1$, Phumudzo Rabambi$^1$, Randle Rabe$^1$ and Sanjaye Ramgoolam$^{1,2}$}
 \email{robert@neo.phys.wits.ac.za, 457990@students.wits.ac.za, randlerabe@gmail.com, s.ramgoolam@qmul.ac.uk}
\affiliation{
School of Physics and Mandelstam Institute for Theoretical Physics$^1$,\\
University of Witwatersrand, Wits, 2050, South Africa
}
\affiliation{Centre for Research in String Theory and School  of Physics and Astronomy$^2$,\\
Queen Mary University of London, Mile End Road, London E1 4NS UK}

\date{\today}

\begin{abstract}
We develop general counting formulae for primary fields in free four dimensional (4D)  scalar  conformal field theory (CFT). 
Using a duality map between primary operators in scalar field theory  and multi-variable polynomial functions
subject to differential constraints, we identify a sector of holomorphic primary fields corresponding to polynomial functions on a class of 
permutation orbifolds. These orbifolds have palindromic Hilbert series, which indicates they are Calabi-Yau.  We construct the top-dimensional holomorphic form expected from the Calabi-Yau property.  This sector includes and extends previous constructions of infinite families of  primary fields.  We sketch the generalization of these results to free 4D vector and matrix CFTs.

\end{abstract}

\pacs{Valid PACS appear here}

\maketitle

\section{\label{sec:Introduction} Introduction  }

In \cite{Koch:2014nka} we started a program of describing the discrete combinatoric data of four dimensional conformal field theories (CFT4)  using the framework of $SO(4,2)$ invariant  2D topological field theory (TFT2).  TFT2 associates state spaces to circles and the operator product expansion of the 4D CFT determines amplitudes for 3-holed spheres. We described  how the associativity conditions of 2D TFT are satisfied by  the correlators of free scalar CFT4. We initated the investigation of   $SO(4,2)$ invariant TFT2 as an approach to 
perturbative field theory in \cite{INTINT}, making contact with the equivariant interpretation of conformal  Feynman integrals in mathematical work \cite{FL}. In this paper we return to free scalar CFT4 and develop the concrete counting and construction of primary fields, which gives the decomposition of the state space of the 2D TFT in terms of  $SO(4,2)$ representations.  
 Another motivation of this paper is that elegant arguments have been found which relate explicit information on the combinatorics of  primary fields and OPE coefficients of  free CFT4  to observables in the epsilon expansion  \cite{Rychkov:2015naa,Basu:2015gpa,Ghosh:2015opa,Raju:2015fza,Nii:2016lpa}. 

Using a duality between primary fields and multi-variable polynomials, we map the problem of constructing primary fields into a many-body quantum mechanics problem, 
equivalently a problem of solving a system of linear constraints for functions 
 $ \Psi ( x^I_{\mu})$. 
 This system of equations is given in equation (\ref{constraints}). 
One of these constraints is a harmonicity condition in $( \mathbb{R}^4 )^{ \times n} $. A large class of solutions is obtained by choosing a complex structure $ ( z, w )$ on 
$ \mathbb{ R}^4 = \mC^2 $ and restricting to holomorphic solutions. The special case where the functions depend on a single complex variable makes contact with previously available explicit construction of primary fields in the literature. 
The reduction of the second order constraint to first order holomorphic conditions 
has the intriguing consequence that these primary fields have a closed ring structure. The associated generating functions have palindromicity poperties due to the fact that the primary fields correspond to functions on the Calabi-Yau orbifolds 
 \bea 
 { ( \mC^2)^{ n  }  /  ( \mC^2 \times S_n )   }  
 \eea 
 which can also be written as 
 \bea
  (  \mC^n / \mC \times \mC^n / \mC ) / S_n 
\eea
where $n$ is the number of elementary fields $\phi$. A  generalization  of our discussion to the $O(N)$ vector model  shows 
 that the holomorphic singlet primary fields correspond to functions 
on the Calabi-Yau orbifold
\begin{equation} 
(\mC^{2} )^{ 2n } /( \mC^2 \times S_{n}[S_2] )   = ( \mC^{2n} / \mC \times \mC^{ 2n } / \mC ) / S_{n} [ S_2]   
\end{equation} 
where $ S_n[S_2]$ is a wreath product subgroup of $ S_{2n}$. For the 
the matrix model in which $\phi$ transforms in the adjoint of $U(N)$, we find holomorphic primaries corresponding to 
polynomial functions on 
\begin{equation} 
( ( \mC^2 )^{  n } \times S_n )  / (  \mC^2 \times S_n )  
\end{equation} 
which can also be written as 
\begin{equation} 
(\mC^n/\mC \times \mC^n / \mC\times S_n) / S_n   
\end{equation}

\section{\label{sec:PolyRep} Multi-variable Polynomial (many-body) representation  of $SO(4,2)$ }

In radial quantization, the scalar field has a mode expansion given by
\bea\label{RQradmodexp}  
\phi ( x_{ \mu } ) = \sum_{ l =0  }^{ \infty } \sum_{ m \in V_l }   a^{\dagger}_{ l , m } Y_{ l , m } ( x )  
 + \sum_{ l = 0 }^{ \infty } \sum_{ m \in V_l } a_{ l , m } |x|^{-2} Y_{ l ,m } ( x' ) \cr 
\eea
$V_l$ is the representation of $SO(4)$ corresponding to symmetric traceless tensors of rank $l$.  The index $m$ runs over a basis for this vector space. 
Acting on the vacuum state $|0\rangle$ (which is, by definition, annihilated by the $a_{l,m}$) with a local operator $\partial_{\mu_1}\cdots \partial_{\mu_k}\phi$ and taking the limit $ x \rightarrow 0$, we get a state. Taking the dual of this state and pairing with $\phi(x)|0\rangle$ we get a polynomial. 
Thus there is a map 
\begin{eqnarray}
   \partial_{\mu_1}\cdots \partial_{\mu_k}\phi \leftrightarrow P_{\mu_1}\cdots P_{\mu_k}\cdot 1
\end{eqnarray} 
where \cite{Koch:2014nka}
\bea 
P_{ \mu} = x^2 \partial_{ \mu} - 2 x_\mu x . \partial - 2 x_{\mu} 
\eea
The scalar field itself maps to  $1$. The free field in (\ref{RQradmodexp}) satisfies 
the equation of motion $ \partial_{ \mu} \partial_{ \mu} \phi  = 0$. Correspondingly $ P_{\mu} P_{\mu} 
= x^4 \partial^2 $ annihilates $1$. 
 When considering operators constructed 
using $n$ fields, we have a representation of the conformal group on 
polynomials in variables $x^I_\mu$ where $ I $ ranges from $1$ to $n$. The  
generators for special conformal transformations and translations are \cite{Koch:2014nka}
\begin{eqnarray}
   K_\mu = \sum_{I=1}^n {\partial\over\partial x^I_\mu}
\end{eqnarray}
\begin{eqnarray}
P_\mu =\sum_{I=1}^n \left(x^{I\rho}x^I_\rho {\partial\over x^I_\mu} -2x^I_\mu x^I_\rho {\partial\over x^I_\rho}
-2x^I_\mu\right)
\end{eqnarray}
The remaining generators are determined by the $so(4,2)$ algebra.
The $x^I_\mu$ can be considered as the coordinates of $n$ particles.
The construction of primaries using $n$ copies of the elementary field $ \phi$ is therefore mapped to a many-body quantum mechanics problem with $n$ particles. 

Tracelessness can be implemented \cite{Dobrev:1975ru,Costa:2011mg} using 
variables $z\cdot x^I = z^\mu x^I_\mu$ with null $z^\mu$: $z^\mu z_\mu =0$.
Any polynomial in $z\cdot x^I$ gives a traceless symmetric polynomial in $x^I_\mu$ after the $z^\mu$s
are stripped away.
The translation between polynomials and operators is
\begin{eqnarray}
  (z\cdot\partial )^k\phi \leftrightarrow (-1)^k 2^k k! (z\cdot x)^k\label{polytranslate}
\end{eqnarray}
This construction is not general: there are primaries that are not symmetric in
their indices and so can't be represented as a polynomial in $z\cdot x$.
For the general discussion, introduce projectors from symmetric tensors to traceless symmetric tensors. 
For example, for tensors of rank 2 and 3 we have
\begin{eqnarray} 
S_{\mu\nu}^{ \alpha \beta} & = &  \delta^{\alpha }_{\mu} \delta^{\beta}_{\nu} - {1\over 4}\delta_{\mu\nu} \delta^{\alpha\beta}\cr 
S_{\mu\nu\rho}^{\alpha\beta\gamma}& = &\delta^{\alpha}_{\mu}\delta^{\beta}_{\nu}\delta^{\gamma}_{\rho} 
   - {1\over 6} (\delta_{\mu\nu} \delta^{\alpha\beta}\delta_\rho^{\gamma}
+\delta_{\mu\rho}\delta^{\alpha\gamma}\delta_\nu^\beta
+ \delta_\mu^\alpha \delta^{ \beta \gamma}\delta_{\nu\rho}) 
\cr
&&
\end{eqnarray}
We recognize that these are projectors in the Brauer algebra of tensor space operators which commute with $SO(4)$ \cite{GoodWall}
\begin{eqnarray} 
S^{(2)} & = &  1 - {C_{12}\over 4}\cr  
S^{(3)} & = & 1 - {1\over 6} \left (C_{12} + C_{13} + C_{23} \right ) 
\end{eqnarray}
They satisfy 
\begin{eqnarray}
\label{prod}  
(S^{(n)} )^2 P_{n}=S^{(n)} P_{n} 
\end{eqnarray}
where
\begin{eqnarray} 
P_{ n} = { 1 \over n! } \sum_{ \sigma \in S_n } \sigma 
\end{eqnarray}
The projector property along with the property that they start with $1$ completely determines these elements of the Brauer algebra. 
In general
\begin{eqnarray} 
P_{\mu_1}\cdots P_{\mu_n}\cdot  1= (-1)^n {2^n n!} (S^{(n)} )_{\mu_1\cdots\mu_n}^{\nu_1\cdots\nu_n}x_{\nu_1} \cdots x_{\nu_n} \nonumber
\end{eqnarray}
The multiplication (\ref{prod})  is in the Brauer algebra, where loops are assigned the value of $4$. 
The factor on the RHS above is obtained by deriving an obvious recursion formula. 
Note that the term $x^2 \partial_{\mu}$ in $P_\mu$ does not raise the rank of the tensor. 
The other two terms contribute to the recursion. 

States at dimension $n+k$ in $V^{\otimes n}$ correspond to polynomials in $x^{I}_{\mu} $ of degree $k$. 
Primaries at dimension $n+k$ are degree $k$ polynomials $\Psi(x^I_{\mu})$  with the conditions
\begin{eqnarray}\label{constraints}  
&&K_{\mu}\Psi( x_{\mu}^{ I } ) = \sum_{I}\frac{\partial}{\partial x^I_{\mu}}  \Psi( x_{\mu}^{ I } ) =0 \cr 
&&{\cal L}_{I}  \Psi ( x_{\mu}^{ I } )= \sum_{\mu} \frac{\partial}{\partial x^I_{\mu}}
\frac{\partial}{\partial x^I_{\mu}}\Psi ( x_{\mu}^{ I } ) =0  \cr
&&   \Psi ( x_{\mu}^{ I } ) = \Psi ( x_{ \mu}^{ \sigma (I) } )
\end{eqnarray}
The first condition says the special conformal generators annihilate a primary operator.
The second condition implements the free scalar equation of motion. 
The last condition imposes $S_n$ invariance,  to implement bosonic statistics of the scalar field. 

We find it useful to employ the complex coordinates
\begin{eqnarray}
  z=x_1+ix_2\qquad w=x_3+ix_4\cr
 \bar z=x_1-ix_2\qquad \bar w=x_3-ix_4
\end{eqnarray}
which have the following $(j^3_L,j^3_R)$ charge assignments
\begin{eqnarray}
&&z\leftrightarrow ({1\over 2},{1\over 2})\qquad
\bar z\leftrightarrow (-{1\over 2},-{1\over 2})\cr
&&w\leftrightarrow ({1\over 2},-{1\over 2})\qquad
\bar w\leftrightarrow (-{1\over 2},{1\over 2})
\end{eqnarray}

This amounts to choosing an isomorphism between $ \mathbb{R}^4 $ and 
$ \mathbb{C} \times \mathbb{C}$. 
We will construct a class of primaries corresponding to 
holomorphic polynomial functions on 
\begin{eqnarray} 
{  \mathbb{C }^{2 n} / (  \mathbb{C}^2   \times  S_n )  } 
\end{eqnarray} 

\section{\label{sec:Counting} Counting with $SO(4,2)$ characters }

The number $N_{[\Delta,j_1,j_2]}$ of primary operators, of dimension $\Delta$ and spin
$(j_1,j_2)$ built out of $n$ scalar fields $\phi$ is obtained by expanding the generating function
\begin{eqnarray} 
   G_n (s,x,y)=\sum_{m,j_1,j_2}N_{[m,j_1,j_2]}s^m x^{j_1} y^{j_2}
\end{eqnarray}
The generating function is given by (take $n\ge 3$ to avoid complications associated to
null states)
\begin{eqnarray}
  G_n (s,x,y)&&=\left[ (1-{1\over x})(1-{1\over y})Z_n (s,x,y)
(1-s\sqrt{xy})\right.\cr
&&\left. (1-s\sqrt{x\over y})
(1-s\sqrt{y\over x})(1-{s\over\sqrt{xy}})\right]_{\ge}
\end{eqnarray}
where $Z_n (s,x,y)$ is defined by
\begin{eqnarray}
\prod_{q=0}^\infty\prod_{a=-{q\over 2}}^{q\over 2}\prod_{b=-{q\over 2}}^{q\over 2}
{1\over 1-t s^{q+1}x^a y^b}=\sum_{n=0}^\infty t^n Z_n (s,x,y)
\end{eqnarray}
This is obtained by constructing the character for the symmetric product of $n$ copies of the
representation of the scalar field, and decomposing into $SO(4,2)$ irreps\cite{Newton:2008au,long}.

We can specialize this counting formula.
Consider the leading twist fields, with $[\Delta,j_1,j_2]=[n+q,{q\over 2},{q\over 2}]$.
This is a complete spin multiplet. The highest spin primary corresponds to a polynomial in $z$.
For counting these primaries, the general formulas given above reduce to
\begin{eqnarray}
  G_n^{z} (s,x,y)=\left[ Z^{z}_n (s,x,y)
(1-s\sqrt{xy})\right]
\end{eqnarray}
where
\begin{eqnarray}
\prod_{q=0}^\infty
{1\over 1-t s^{q+1}x^{q\over 2} y^{q\over 2}}=\sum_{n=0}^\infty t^n Z^{z}_n (s,x,y)
\end{eqnarray}
Using the simplified formulas we have
\begin{eqnarray}
G^{\rm max}_n (s)=
{s^n\over (1-s^2)(1-s^3)\cdots (1-s^n)}\label{Polyinz}
\end{eqnarray}
Note the close connection to multiplicities of $V^{SL(2)}_{\Lambda =n+k}\otimes V^{S_n}_{[n]}$, which is 
the coefficient of $q^k$ in
\begin{eqnarray}
\prod_{i=2}^n {1\over 1-q^i}
\end{eqnarray}
The result (\ref{Polyinz}) was also recently obtained in \cite{Roumpedakis:2016qcg}.

A more general counting involves polynomials of $ z_{\mu}^I  $ and $w_{\mu}^I $. 
We denite these as expremal primaries, since they have $ ( s , j_L , j_R ) = ( n + q , { q \over 2 } , j_R )$.  In this case
\begin{eqnarray}
G^{z,w}_n(s,x,y)=\left[\left(1-{1\over y}\right)(1-s\sqrt{xy})\right.\cr
\left. (1-s\sqrt{x/y})Z_n(s,x,y)\right]_{\ge} \label{zwcount}
\end{eqnarray}
where
\begin{eqnarray} 
\prod_{q=0}^{\infty}\prod_{m=0}^{q}{1\over (1- t s^{q+1}x^{q/2}y^{m-q/2})}
=\sum_{n=0}^{\infty}t^n Z^{z,w}_{n} (s,x,y) \nonumber
\end{eqnarray}
As explained in more detail in Section 4, 
\bea\label{ZnZL1} 
&& Z^{z,w}_{n} (s,x,y) = s^n\sum_{ \Lambda_1} 
Z_{\rm{ SH } }  ( s \sqrt { x y  } , \Lambda_1 ) Z_{SH  }  (s \sqrt{ x\over  y}   , \Lambda_1 ) \cr &&
\eea
where $ \Lambda_1 $ is a partition of $n$, and $ Z_{SH} ( q ) $ is given in 
(\ref{ZSH}). 
Using these formulae, one finds, for $n=3$
\begin{eqnarray}\label{Z3sxy}  
&&Z^{z,w}_3 = {s^3\left(s^6x^3+s^4x^2+s^2x+1+s^3x^{3\over 2}\left(\sqrt{y}+{1\over\sqrt{y}}\right)\right)\over
(1-s^2 xy)(1-s^3(xy)^{3\over 2})(1-s^2{x\over y})(1-{s^3x^{3\over 2}\over y^{3\over 2}})}\cr
&&G^{z,w}_3={s^3 (1+s^5 x^{5\over 2}y^{3\over 2})\over (1-s^4 x^2)(1-s^3\sqrt{x^3y^3})(1-s^2 xy)}
\label{countnis3}
\end{eqnarray}

Computing $ Z_4 ( s, x, y ) $ in the same way we find 
\begin{eqnarray}
G^{z,w}_4={s^4 R(s,x,y)\over 
D(s,x,y)} 
\end{eqnarray}
where
\begin{eqnarray}
R(s,x,y)&=&1+s^5 x^{5\over 2} \big(\sqrt{y}+s^3 x^{3\over 2} y+s^5 x^{5\over 2} y+y^3-s^6 x^3 y^{5\over 2}\cr
&&-s^8 x^4 y^{5\over 2}-s^{16} x^8 y^{7\over 2}-s^{11} x^{11\over 2} y^2 \left(1+y\right)\cr
&&+s^7 x^{7\over 2} \left(1-y^2\right)+s^4 x^2 y^{3\over 2} \left(1-y^2\right)\cr
&&+s^2 x\sqrt{y} \left(1+y^2\right)
-s^9 x^{9\over 2} y \left(1+y^2\right)\cr
&&-s^{10} x^5 y^{3\over 2} \left(1+y-y^2\right)-s \sqrt{x} \left(1-y-y^2\right)\big)\nonumber
\end{eqnarray}
\begin{eqnarray}
D(s,x,y)&=&(1-s^2 x y)(1-s^3 x^{3\over 2} y^{3\over 2})(1-s^4 x^2 y^2)\cr
&& (1-s^4 x^2)\left(1-s^6 x^3\right) \left(1-s^8 x^4\right)\nonumber
\end{eqnarray}
Similar constructions with the pairs $ ( z , \bar w ), ( \bar z , w ) , ( \bar z , \bar w )$ are possible. 

\section{\label{sec:Construction} Counting and Construction with symmetric groups }

The counting formulas derived in section \ref{sec:Counting} can be used to construct families of primary operators.
The coordinates $x^I_\mu$, $I=1,...,n$ admit a natural action of $S_n$.
To satisfy the first of (\ref{constraints}), build $n-1$ translation invariant ``relative coordinates''
given by the successive differences $X_{\mu}^{(a)} = x_{\mu}^{a} - x_{\mu}^{ a+1}  $. Using   the complex coordinates $ z , w  $ on $ \mathbb{R}^4 = \mC^2 $, we have $ ( z_{ I } , w_I ) $ 
on $ (\mathbb{R}^4)^n = ( \mC^2 )^n $. These differences span the $S_n$ irrep labeled by hook Young diagram with row lengths $ [ n-1,1]$.  A more convenient basis which connects with 
Young's orthonormal representation is useful for computations (see \cite{BHR} for this basis).  Using complex variables we have  $Z^{ (a)} , \bar Z^{(a)}$, $Z^{(a)}$ and $\bar W^{(a)}$, which each transform in the irrep $[n-1,1] \equiv V_H$. 
Products $Z^{(a_1)}Z^{(a_2)}\cdots Z^{(a_k)}$ are in the $V_H^{\otimes k}$ tensor product representation of $S_n$.
Any polynomial in the hook variables automatically obeys the first two constraints of (\ref{constraints}). This follows since the Laplacian in the second equation is 
\bea 
\left ( { \partial^2 \over \partial z^I \partial \bar z^I }  + { \partial^2 \over \partial w^I \partial \bar w^I }  \right ) \Psi = 0 
\eea
The only thing left is to project to the $S_n$ invariant subspace of $V_{H}^{\otimes k}$.  The matrices 
representing the $k$-fold tensor product are
\begin{eqnarray}
&& (  \Gamma_k^{n}  (\sigma) )_{a_1, \cdots  , a_k ; b_1  , \cdots , b_k} 
=\Gamma_{a_1, b_1 }  (\sigma)\cdots\Gamma_{a_k , b_k }(\sigma) \nonumber
\end{eqnarray}
where $ \Gamma_{a,b} ( \sigma ) $ are  matrices representing $S_n$
in  an orthogonal basis of $ [n-1,1]$. 
We can project to the invariants by averaging over the group 
\begin{eqnarray}
   P_{a_1 a_2\cdots a_k}={ 1 \over n! } \sum_{\sigma\in S_n } ( \Gamma_k (\sigma))_{a_1 a_2\cdots a_k,b_1 b_2\cdots b_k}
Z^{(b_1)}\cdots Z^{(b_k)}\nonumber
\end{eqnarray}
The above expression gives $\sum_i \hat{n}_i P_i(z)$ where $\hat{n}_i$ are unit vectors and $P_i(z_1 , \cdots , z_n )$ are the polynomials we want.

By considering all possible degrees  $k \in \{ 0 , 1, 2 , \cdots \}$ we have a ring. 
These primaries have a ring structure, since they obey a stronger linear version of 
the Laplacian condition, which means that a product of solutions is also a solution to 
the constraints. 
The counting formula (\ref{Polyinz})  gives the Hilbert series for holomorphic functions on $ ( \mC^n / \mC ) / S_n $. The quotient by $\mC$ is effected by the first of (\ref{constraints}) which sets the centre of mass momentum of the
many body wavefunction to zero. 
The  orbifold by  $S_n$ is the symmetry condition in (\ref{constraints}). Using properties of Hilbert series, it follows that the ring at hand has $n-1$ generators,
whose form is outlined in the Appendix \ref{LTGens}.

The construction is easily extended to polynomials of holomorphic coordinates $z^I$ and $w^I$. Use hook variables 
$Z^{(a)} , W^{(a)} $. The products $ Z^{(a_1)} \cdots Z^{(a_k)} W^{(a_{k_1})} \cdots W^{(a_{k+l})}$ 
belong to a subspace of  the  representation $ V_{H}^{ \otimes k } \otimes  V_H^{ \otimes l } $ of $S_n$, which we will characterize in terms of representation theory. 
Consider the expansions in terms of $ S_n \times S_k$ irreps  
\begin{eqnarray}\label{exphooks}
&& V_H^{ \otimes k } = \bigoplus_{ \Lambda_1 \vdash n , \Lambda_2 \vdash k } V_{ \Lambda_1}^{  (S_n)} \otimes V_{ \Lambda_2}^{ (S_{k} ) } \otimes V^{ Com ( S_n \times S_k )}_{ \Lambda_1 , \Lambda_2} \cr 
&& V_H^{ \otimes l  } = \bigoplus_{ \Lambda_3 \vdash n , \Lambda_4 \vdash l  }  V_{ \Lambda_3}^{  (S_n)} \otimes V_{ \Lambda_4}^{ (S_{l} ) } \otimes V^{ Com ( S_n \times S_l )}_{ \Lambda_3 , \Lambda_4} 
\end{eqnarray}
Multiplicities are given by dimensions of irreps of the commutants $ Com ( S_n \times S_k )$ in $ V_H^{ \otimes k }$.  Since the $Z$ and $W$ variables are commuting, 
the monomials belong to the trivial irreps $ \Lambda_2 = [k] \otimes \Lambda_4 =  [l]$ of $ S_k \times S_l$. To satisfy the third constraint, project to $S_n$ invariants in $V_{H}^{\otimes k}\otimes V_H^{\otimes l}$. This constrains $\Lambda_3=\Lambda_1$.  
So the number of $ S_k \times S_l \times S_n$ invariants is 
\begin{eqnarray}
\sum_{\Lambda_1\vdash n} {\rm Mult}(\Lambda_1,[k];S_n\times S_k) ~~ {\rm Mult}(\Lambda_1,[l];S_n\times S_l) 
\nonumber
\end{eqnarray}
The expansions (\ref{exphooks}) are explained further and used in the construction of BPS states in \cite{BHR}. 
The generating functions for these multiplicities are derived in \cite{BHR}. 
$Mult (\Lambda_1,[k];S_n\times S_k) \equiv Z_{ SH}^k $ is the coefficient of $q^k$ in 
\begin{eqnarray}\label{ZSH}
Z_{SH} (q;\Lambda_1)&=&( 1- q ) ~  q^{\sum_{i} c_i(c_i-1)\over 2} ~ \prod_b{1\over (1-q^{h_b})}\cr
&=&\sum_{k}q^k Z_{SH}^k (\Lambda_1) 
\end{eqnarray}
Here $c_i$ is the length of the $i$'th column in $ \Lambda_1$, $b$ runs over boxes in the Young diagram $\Lambda_1$ 
and $h_b$ is the hook length of the box $b$. 
Thus, for the number of primaries constructed from $ z_i , w_i $ we get 
\begin{eqnarray} 
\sum_{ \Lambda_1 \vdash n } Z_{SH }^k  ( \Lambda_1 )
Z_{SH }^l ( \Lambda_1 )
\end{eqnarray}
These  are primaries of weight $n+k+l$, with $(J_3^L,J_3^R)=({k+l\over 2},{k-l\over 2})$. 
We can also show directly that $Z_n ( s , x, y ) $ in   (\ref{zwcount})
is a sum over irreps $ \Lambda_1 $ of $S_n$ as above. Thus
\bea 
Z_n ( s , x, y ) = \sum_{ \Lambda_1 \vdash n }\sum_{k,l} Z_{SH }^k  ( \Lambda_1 )
Z_{SH }^l ( \Lambda_1 )s^{n+k+l}x^{k+l\over 2} y^{k-l\over 2}\cr
= s^n  \sum_{ \Lambda_1 \vdash n } Z_{SH} ( q = s \sqrt{ xy} , \Lambda_1 ) Z_{SH} ( q = s \sqrt{ x/y} , \Lambda_1 ) \cr 
\eea
Using the generating function (\ref{ZSH}), we get  the rational expressions for $ Z_3 ( s , x, y ) , Z_4 ( s, x, y ) $ used in section \ref{sec:Counting}, 
by explicitly doing the sum over $ \Lambda_1$. 

This structure in the counting problem provides an explicit construction formula.
First, decompose the $z$ and $w$ polynomials into definite $S_n$ irreps.
The projector onto irrep $r$ from the tensor product of $k$ copies of the hook is
\begin{eqnarray}
 P^{ \Lambda_1}_{a_1 \cdots a_k,b_1 \cdots b_k}={1\over n!}\sum_{\sigma\in S_n}\chi_{ \Lambda_1} (\sigma)
(\Gamma_k^n  (\sigma))_{a_1 \cdots a_k,b_1 \cdots b_k}\nonumber
\end{eqnarray}
We also need the projection onto the symmetric irrep
\begin{eqnarray}
P_{a_1\cdots a_n,b_1\cdots b_n}={1\over n!}\sum_{\sigma\in S_n}
(\Gamma_k^n (\sigma))_{a_1 \cdots a_n\, ,\, b_1 \cdots b_n}
\end{eqnarray}
Using these two projectors, the polynomials constructed using two holomorphic variables are
\begin{eqnarray}
\sum_A P_A( \vec z, \vec w)\hat n^A_{e_1\cdots e_{k+l}}
=P_{e_1\cdots e_{k+l},a_1\cdots a_k c_1\cdots c_l}\cr
\times P^r_{a_1 \cdots a_k,b_1 \cdots b_k}P^r_{c_1 \cdots c_l,d_1 \cdots d_l}
Z^{(b_1)}\cdots Z^{(b_k)}W^{(d_1)}\cdots W^{(d_l)}
\nonumber
\end{eqnarray}
where $\hat{n}^A$ are unit vectors and $P_A(z^I ,w^I )$ are the polynomials corresponding to primary fields. These polynomials satisfy all the conditions 
in (\ref{constraints}). They satisfy  stronger linear equations 
\bea 
\partial_{ \bar w^J } P_A(  z^I  ,  w^I   )= 0 ~~ ; ~~ \partial_{ \bar z^J } P_A(z^I ,w^I )= 0
\eea
which imply the Laplacian conditions. As a result,  taking all possible $ k , l $, we 
have a space of solutions to the constraints which forms a ring due to the Leibniz rule for 
 products of functions. This is the polynomial ring of holomorphic functions 
  for 
  \bea\label{CYorbs}  
(\mC^2 )^n / (\mC^2 \times S_n ) 
  \eea

Using generalities about Hilbert series for algebraic varieties (see \cite{GeomAP,SQCDAd} 
for applications in the context of moduli spaces of SUSY gauge theories), we see from 
 (\ref{countnis3}) that for $n=3$ the polynomials $P_A(z,w)$ are a finitely generated polynomial ring with $3$ generators. The explicit constructions described above allow us to identify the 
 generators ($z_{ij}\equiv z_i-z_j$)
\begin{eqnarray}
(z_{12})^{2k}+(z_{13})^{2k}+(z_{23})^{2k} &&\leftrightarrow\quad (s^2 xy)^k\cr\cr
(z_{13}+z_{23})^k (z_{31}+z_{21})^k (z_{12}+z_{32})^k
 &&\leftrightarrow\quad (s^3\sqrt{x^3y^3})^k\cr\cr
\left|
\begin{matrix}
w_1 &w_2 &w_3\cr z_1 &z_2 &z_3\cr 1 &1 &1
\end{matrix}\right|^{2k}
&& \leftrightarrow \quad (s^4 x^2)^k
\end{eqnarray}
This is explained in more detail in the forthcoming paper \cite{long}. 

The Hilbert series associated to the counting of primary fields ensures a palindromic property 
of the numerators. This can be verified for $ Z_{3 } ( s, x, y) , Z_4 ( s, x, y )$. 
A general property of the numerators
\bea 
Q_n ( s , x , y )   = \sum_{ k =0 }^D a_k ( x, y ) s^k 
\eea
is that $ a_{ D-k}  (  x , y ) = a_{ k} ( x , y  )$. A direct proof using the combinatoric expressions 
like equation (\ref{ZnZL1})  in terms of symmetric group representation theory data, is given in \cite{long}.
The theorem of Stanley\cite{Stanley} suggests that these orbifolds are Calabi-Yau. This can be explicitly
demonstrated by constructing the top form and verifying that it is nowhere vanishing\cite{long}.

The above argument starting from counting to motivate a construction of the primary operators and then an
associated Calabi-Yau geometry goes through when the single scalar is generalized to the $O(N)$ vector
model and to the free $U(N)$ gauge theory with $\phi$ a matrix in the adjoint.
The relevant geometries are the Calabi-Yau orbifolds
\begin{equation} 
(\mC^{2})^{2n}/(\mC^2 \times S_{n}[S_2])  \label{CY}
\end{equation} 
and
\begin{equation} 
((\mC^2)^n\times S_n) / (\mC^2\times S_n) \label{CYt}  
\end{equation}
respectively.
It is fascinating that non-trivial properties of  the combinatorics of primary fields in free four dimensional 
conformal field theory is related to the  geometry of Calabi-Yau orbifolds (\ref{CYorbs}), (\ref{CY}) and (\ref{CYt}).

\begin{acknowledgments}
This work of RdMK, PR and RR is supported by the South African Research Chairs
Initiative of the Department of Science and Technology and National Research Foundation
as well as funds recieved from the National Institute for Theoretical Physics (NITheP).
SR is supported by the STFC consolidated grant ST/L000415/1 “String Theory, Gauge Theory \& Duality”
and  a Visiting Professorship at the University of the Witwatersrand, funded by a Simons Foundation 
grant held at  the Mandelstam Institute for Theoretical Physics.
\end{acknowledgments}

{\vskip 0.25cm}

\appendix

\section{Appendix: Leading Twist Generators}\label{LTGens}

The counting formula (\ref{Polyinz}) demonstrates that the leading twist primaries form a ring generated by
$n-1$ generators.
These generators are given by constructing the $n-1$ possible independent $S_n$ invariants out of the hook variables,
which are given by\cite{BHR}
\begin{eqnarray}
X_\mu^{(a)}={1\over\sqrt{a(a+1)}}(x^1_\mu+\cdots+x^a_\mu-a x^{a+1}_\mu)\label{hookvars}
\end{eqnarray}
For example, for $n=2$ fields the polynomials are generated by $(z_1-z_2)^2$.
The polynomials corresponding to primaries are
\begin{eqnarray}
(z_1-z_2)^{2k}
\end{eqnarray}
Using (\ref{polytranslate}) it is easy to see that (these primaries vanish if $s$ is odd)
\begin{eqnarray}
O_s &=&(z_1-z_2)^s\cr
&\leftrightarrow& {s!\over 2^s}\sum_{k=0}^s 
{(-1)^k \over (k! (s-k)!)^2}\, \partial_z^{s-k}\phi\, \partial_z^{k}\phi
\end{eqnarray}
reproducing the higher spin currents, given for example in \cite{Giombi:2016hkj}.
For $n=3$ fields the ring of polynomials that correspond to primary operators is generated by
\begin{eqnarray}
(z_1-z_2)^{2}+(z_1-z_3)^{2}+(z_2-z_3)^{2}
\end{eqnarray}
and
\begin{eqnarray}
(z_1+z_2-2z_3) (z_3+z_2-2z_1) (z_1+z_3-2z_2)
\end{eqnarray}

{}

\end{document}